\documentclass[apj]{emulateapj}
\usepackage[colorlinks,linkcolor={blue},citecolor={blue},urlcolor={red}]{hyperref}
\usepackage{epsfig}
\usepackage{graphicx}
\usepackage{natbib}
\usepackage{amsmath}
\usepackage{amsfonts}
\usepackage{amssymb}
\usepackage{xcolor}
\def\mpchi{\,h^{-1}{\rm {Mpc}}}
\def\kpchi{\,h^{-1}{\rm {kpc}}}
\def\kpc{\,{\rm {kpc}}}
\def\kms{\,{\rm {km\, s^{-1}}}}
\def\msun{{\rm M_\odot}}
\def\msunhi{\,h^{-1}{\rm M_\odot}}

\slugcomment{}

\begin{document}

\title{A scaling relation between merger rate of galaxies and their close pair count}
\author{C. Y. JIANG \altaffilmark{1}, 
Y. P. JING\altaffilmark{2,\ref{email}},  
JIAXIN HAN\altaffilmark{3},
}
\altaffiltext{1}{Key Laboratory for Research in Galaxies and Cosmology of
Chinese Academy of Sciences, Shanghai Astronomical Observatory,
Nandan Road 80, Shanghai 200030, China} 
\altaffiltext{2}{Center for Astronomy and Astrophysics, Department of 
Physics and Astronomy, Shanghai Jiao Tong University, Shanghai 200240, China; 
\label{email}\textcolor{blue} {\texttt{ypjing@sjtu.edu.cn}}}
\altaffiltext{3}{Institute for Computational Cosmology, Department of Physics, 
University of Durham, Science Laboratories, South Road, Durham DH1 3LE, UK}

\begin{abstract}

  We study how to measure the galaxy merger rate from the observed
  close pair count. Using a high-resolution N-body/SPH cosmological
  simulation, we find an accurate scaling relation between galaxy pair
  counts and merger rates down to a stellar mass ratio of about 1:30.
  The relation explicitly accounts for the dependence on redshift (or
  time), on pair separation, and on mass of the two galaxies in a
  pair. With this relation, one can easily obtain the mean merger
  timescale for a close pair of galaxies. The use of virial masses,
  instead of stellar masses, is motivated by the fact that the
  dynamical friction time scale is mainly determined by the dark
  matter surrounding central and satellite galaxies. This fact can
  also minimize the error induced by uncertainties in modeling star
  formation in the simulation.  Since the virial mass can be read
  from the well-established relation between the virial masses and the
  stellar masses in observation, our scaling relation can be easily
  applied to observations to obtain the merger rate and merger time
  scale.  For major merger pairs (1:1-1:4) of galaxies above a stellar
  mass of $4\times10^{10}\msunhi$ at $z=0.1$, it takes about $0.31$
  Gyr to merge for pairs within a projected distance of $20\kpchi$
  with stellar mass ratio of 1:1, while the time taken goes up to
  $1.6$ Gyr for mergers with stellar mass ratio of 1:4.  Our results
  indicate that a single timescale usually used in literature is not
  accurate to describe mergers with the stellar mass ratio spanning even
  a narrow range from 1:1 to 1:4.
 
\end{abstract}

\section{Introduction}
\label{sec:intro}

Measuring merger rates of galaxies 
is crucial to studies of their growth history. For 
early-type galaxies, merger rates are decisive in quantifying 
their mass assembly history and size increase process. 
It determines whether the frequency of mergers is appropriate 
for the observed mass/size growth of the early-type galaxies
\citep{nipoti09,nipoti12,chou11,bluck12,cimatti12,newman12,mclure13}. For
late-type galaxies, merger rates are important for understanding the
episodic star formation, such as starbursts, which are induced and
enhanced by gas-rich  mergers \citep[e.g.][]{hopkin10,lopez10,lopez13}.

In observational studies, merger rates are usually measured by counting the
number of close galaxy pairs
\citep[e.g.][]{patton97,carlberg00,lin04,ryan08,berrier12}. For
such a measurement, a timescale has to be adopted (or assumed) for a
close pair of galaxies to merge. The uncertainty
associated with the merger timescale also represents the main
uncertainty of the merger rate. In literature,    
the authors generally assume that galaxies
within some projected distance (e.g., $20\kpchi$) would merge in a
constant timescale (around $0.5$ Gyr) after accounting for the
projection effects \citep{patton00,
  patton02,bell06,depropris07,lin04,lin08,patton08}.
\cite{kitzbichler08} checked these assumptions using semi-analytic
catalogues derived from the Millennium Simulation
\citep{springel05}. They found that the merger timescales in their
results are typically larger by a factor of 2 than the constant
timescale that is usually assumed.  In their semi-analytical model,
they traced the dynamics of satellite galaxies through following the
subhalos that host them.  Once the surrounding subhalo is
tidally disrupted, the orphan galaxy is assumed to merge with its central
galaxy after a dynamical friction time.  It is parametrized as in
\cite{binney87}, but multiplied by a fudge factor 2 \citep{kang05,delucia08},
to account for the underestimation of the dynamical friction times in
\cite{binney87}\citep{boylan08,jiang08,wetzel10}.  
The timescales on which close galaxies within a
certain distance would merge, are therefore dependent on the dynamical
times they adopted. Furthermore, the spatial positions of the orphan
galaxies within the host halos cannot be accurately traced by the dark
matter particles, which would affect the measurement of the count of
close pairs. Moreover, \cite{kitzbichler08} only considered major mergers with
stellar mass ratios larger than 1:4. It is hence restricted to major
merger studies, but not applicable for investigating galaxy growth
through minor mergers, which are currently considered the most
promising mechanism for the size evolution of early compact massive
galaxies \citep{bezanson09,naab09,hilz12,oogi13}.

\cite{lotz08,lotz10a,lotz10b,lotz11} analysed the timescales for
mergers identified both with quantitative morphology and with close
pairs using N-body/SPH simulations. For close pair timescales, they
first studied individual merger pairs for several different projected
separation criteria, seeking the dependence on mass ratio, gas
fraction and orbits. To obtain the cosmologically-averaged merger
timescale, they extracted the distributions of baryonic mass ratio and
gas fraction from three cosmological- scale galaxy evolution
models. Then they computed the average merger timescale by weighting
the individual timescale with these distributions. They found that the
timescale depends mainly on the outmost radius adopted to identify
close pairs for major (1:1-1:4) mergers. However, as in
\cite{kitzbichler08}, they did not consider the case of minor mergers
for close pairs either.  Later, \cite{xu12} combined the results in
\cite{lotz10a} with the mass dependence and redshift dependence in
\cite{kitzbichler08}, giving a description for timescales of major
mergers with projected distance between $5\kpchi$ and $20\kpchi$ at 
redshift $z<1$.

In this work, we study the relation between the galaxy merger rate and
number of close pairs using a cosmological N-body/SPH
simulation. Compared with the work of \cite{kitzbichler08}, we trace
the merger of galaxies with star particles, therefore we do not make
any assumptions on the merger timescale and spatial positions of the
orphan galaxies. We extend the series of work by Lotz et al to minor
merges, and carefully examine the dependence of the merger rate on
redshift and on pair separation.  As the main result, we give a
calibrated relation between the merger rate and the close pair count,
where diverse factors are taken into account, including redshift, 
pair separation, and galaxy masses. This relation is applicable to both 
major and minor merger pair counts obtained in observations, up to a 
redshift of 2.

By passing, we would like to point out that two types of merger
timescales are often used in the literature. One, used widely in
theoretical modeling of galaxy formation, measures the time elapsed
during which a satellite spirals into the central galaxy from the halo
virial boundary \citep[eg., ][]{jiang08,boylan08,wetzel10}.  The
other, often used in observations, is the time duration after which a
close pair of galaxies at a certain separation is expected to merge
together. The timescale worked out in the paper is the latter which
can be used to measure the merger rate with the observed galaxy pair
counts.

\section{Methods}
\label{sec:method}

We use an N-body/SPH simulation run with Gadget2 code 
\citep{springel05} to calibrate the relation between galaxy 
merger rates and close pair counts. 
The simulation is the same as the one used in \cite{jiang08}, which includes 
the implementation of radiative cooling, star formation and 
supernova feedback \citep{sh03}. The cosmological parameters are
$\Omega _{\Lambda}=0.732$, $\Omega _{\rm m}=0.268$, $\Omega _{\rm
b}$=0.044, $\sigma_{\rm 8}$=0.85, and the Hubble constant is $H_{\rm
0}=100h \kms{\rm Mpc}^{-1}$ with $h=0.71$.  The box is $100\mpchi$ on
a side, with $512^{3}$ dark matter particles and $512^{3}$ gas
particles. The resulting mass resolution for dark matter and gas
particles is $4.6 \times 10^8 \msunhi$ and $9.2 \times 10^7 \msunhi$,
respectively.
The softening length of the gravitational force is roughly 
equivalent to a Plummer force softening of $4.5\kpchi$ (comoving).
The simulation starts from $z = 120$,     
with an equal logarithmic scale factor interval
of $\Delta \ln a=0.01$ between two consecutive outputs.

Dark matter halos are identified using the friends-of-friends (FOF)
method, with a linking length of 0.2 times the mean inter-particle
separation. By default, galaxies are 
identified with the friends-of-friends method, in which a 
linking-length of 0.025 times the mean inter-particle 
separation is applied to star particles. 
A galaxy merger is identified if the satellite galaxy and the
central galaxy begin to have the same descendant at one snapshot, and
continue to have the same descendant for the following four snapshots
($\geq$ half of the dynamical time of a halo). We use this criterion to
ensure that the merger is a real one, but not just a close flyby.

As usual, two galaxy samples are defined in searching for 
close pairs. Galaxies in the primary sample are targeted to 
count their close neighbors from the secondary sample.

The volume merger rate can be written as
\begin{equation}
\label{eq:phi1}
\Phi = n_{\rm 1}N_{\rm p}(<r_{\rm p})/T_{\rm mg},  
\end{equation}
where $n_{\rm 1}$ is the number density of the primary sample,
and $N_{\rm p}(<r_{\rm p})$ is the mean number of close companions 
within a distance of $r_{\rm p}$ for each galaxy in the primary 
sample. 
$T_{\rm mg}$ is the average timescale for the satellite galaxies at $r_{\rm p}$
to merge with the central one. Here the primary sample (assumed to 
be central galaxies) and the secondary sample (assumed to be satellite 
galaxies) are exclusive to each other. Satellite-satellite pairs will 
be discussed in section \ref{sec:sat-sat}.

The major mechanism that drives the merging of the two galaxies 
is the dynamical friction, which would exhaust  
the orbital energy of the satellite after a certain timescale. 
Therefore, we tentatively formulate the merging timescale 
$T_{\rm mg}$ in the form of dynamical friction timescale, 
$T_{\rm mg} \propto \frac{v_{\rm c}r_{\rm p}^2}{Gm_{\rm 2}\ln \Lambda}$ 
\citep{binney87}, where $v_{\rm c}$ is the circular velocity 
of the primary halo assuming an isothermal sphere, and $m_{\rm 2}$ 
is the satellite mass, which is supposed to be the total mass of 
the baryon and the dark matter that are bound in the satellite 
at the time of being observed.       
In simulations, it is usually called subhalo mass. In observations, however, 
it is not an easily accessible quantity. We will discuss more detailedly 
and look for appropriate proxies in the next section.
Although the galaxy pairs are close in distance,  
some may have higher orbital energy that is equivalent to a larger 
circular radius than the considered radial extent. This effect has
been absorbed into the timescale $T_{\rm mg}$.

If the satellite sample is statistically large enough, the
influence of the circularity parameter would be averaged out.
Therefore, the term related to the circularity parameter is
not included here.

Considering $v_{\rm c}\approx\sqrt{\frac{Gm_{\rm 1,v}}{r_{\rm 1,v}}}$ 
in the primary halo, 
where the virial mass $m_{\rm 1,v}$ and virial radius $r_{\rm 1,v}$ 
are defined in terms of $200$ times the critical density of 
the universe, we have
\begin{equation}
\label{eq:Tmg}
T_{\rm mg} \propto \frac{{m_{\rm 1,v}}^{1/2}{r_{\rm p}}^2}{G^{1/2}m_{\rm 2} \ln \Lambda {r_{\rm 1,v}}^{1/2}}. \\
\end{equation}
The Coulomb logarithm can be written as 
$\ln\Lambda \approx \ln\frac{b_{\rm max}}{b_{\rm min}}$,      
where $b_{\rm max}$ and $b_{\rm min}$ are the maximum and 
minimum impact parameters, respectively.
When the satellite is initially outside the primary halo, 
and $b_{\rm max}$ approximates to the virial radius, we have 
$\ln\Lambda \approx \ln\frac{m_{\rm 1,v}}{m_{2,v}}$,
if we take the virial velocity of the host halo as the initial
velocity. Here $m_{1,v}$ and $m_{2,v}$ are the virial masses of 
the primary halo and the satellite halo, respectively, 
before the satellite enters the primary halo.  
In \cite{jiang08}, they found that an alternative expression 
$\ln(1+\frac{m_{\rm 1,v}}{m_{2,v}})$ for the Coulomb logarithm 
performs better in explaining the time interval for the satellite 
between the time of entering the virial radius and merging 
with the central galaxy, mainly for the case of $m_{\rm 1,v}/m_{2,v} \le 4$. 
However, for galaxy pairs in smaller distance, where satellites are well 
within the virial radius, both $b_{\rm max}$ and $b_{\rm min}$
are very uncertain \citep{hashimoto03,Penarrubia04,just11}. 
We therefore do not explore the explicit influence of $\ln\Lambda$, 
while tentatively integrating it 
into a constant coefficient in the following investigation.

Replacing $T_{\rm mg}$ in Equation (\ref{eq:phi1}) with Equation
(\ref{eq:Tmg}), we obtain
\begin{equation}
\label{eq:phi}
\Phi = A_{\ast} \frac{G^{1/2}m_{ 2} {r_{\rm 1,v}}^{1/2}n_{\rm 1}N_{\rm p}(<r_{\rm p})}{{m_{\rm 1,v}}^{1/2}{r_{\rm p}}^{2}}. \\
\end{equation}
If Equation (\ref{eq:phi}) holds, the coefficient $A_{\ast}$ 
would be a constant, regardless of galaxy masses and redshift.
We will verify this relation by checking on the reciprocal of $A_{\ast}$, 
\begin{equation}
\label{eq:invB}
\frac{1}{A_{\ast}} = \frac{1}{\Phi} \frac{G^{1/2}m_{ 2} {r_{\rm 1,v}}^{1/2}n_{\rm 1}N_{\rm p}(<r_{\rm p})}{{m_{\rm 1,v}}^{1/2}{r_{\rm p}}^{2}},  \\
\end{equation}
which would also 
be a constant if the connection between merger rates and close 
pair counts can be established by Equation (\ref{eq:phi}).

\section{Results}

Equation (\ref{eq:invB}) is obtained in the framework of dynamical 
friction, by assuming all merging pairs are central-satellite pairs. 
For satellite-satellite pairs that were central-satellite pairs 
before entering the primary halo, they may also merge. 
\cite{jiang10} showed that, this kind of merger is a similar 
process to that for central-satellite mergers. We will also  
examine if the relation formulated for central-satellite pairs 
applies to satellite-satellite pairs.

\subsection{Central-Satellite Pairs}
\label{sec:cen-sat}
We know that one ruling factor of the dynamical process is the total bound 
mass of the satellite, including the baryonic mass and and the
surrounding dark matter mass. As shown in \cite{jiang08}, the
merging timescale between the time of entering the virial radius and
the final coalescence (type 1 in \S 1) can be established 
through the virial mass of the satellite. 
However, the total satellite mass varies at different
stages of the merger. At the earlier stage when the satellite just
enters the host halo, the dark matter mass dominates the satellite
mass, and thus the merging timescale.  While at the later phase of the
merging, the dark matter is mostly stripped off, and the stellar mass
of the satellite determines when they finally coalesce. Therefore, at
a middle stage when the two galaxies are $\le 100\kpchi$ away, the
satellite has partly lost its dark matter. The satellite mass $m_{2}$
would be difficult to determine, making it the most uncertain quantity
in Equation (\ref{eq:invB}). In this subsection, we will find a mass proxy 
which is observable and can be used for the relation (\ref{eq:invB}).  

\subsubsection{Adopting Stellar Masses of Satellites}

\begin{figure}
\begin{center}
\plotone{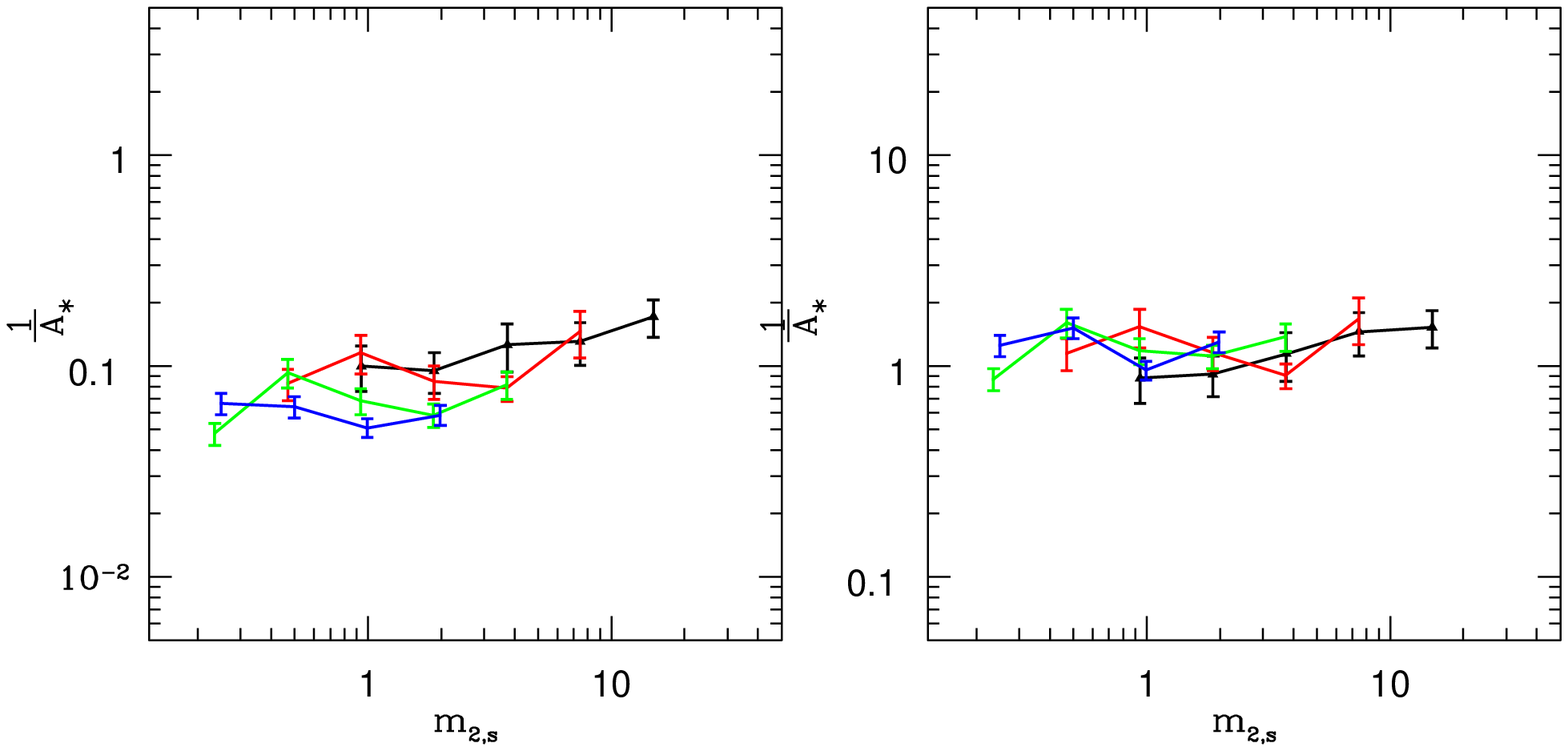}
\caption{\label{fig:msat}
Left: $\frac{1}{A_{\ast}}$ obtained according to Equation (\ref{eq:invB}) 
when the stellar mass $m_{2,s}$ is used for the satellite mass. 
Only central-satellite pairs are included. Four colored lines show 
samples with central stellar masses of $30.0$ (black), $15.0$ (red), 
$7.5$ (green) and $4.0$ (blue), respectively, in unit of $10^{10}\msunhi$.
The outer radius corresponds to $100/(1+z)\kpchi$. 
Error bars show the poisson fluctuations of the merging pairs.
Right: Similar to the left panel, except that the total bound mass of the 
satellite is used in Equation (\ref{eq:invB}). 
}
\end{center}
\end{figure}

 As a first try and as some previous works, we use
  the stellar mass $m_{\rm 2,s}$ to replace the satellite mass $m_{\rm
    2}$ in Equation (\ref{eq:invB}). The advantage to adopt the
  stellar mass is that it can be measured
  in observations.
We set the radial extent $r_{\rm p}$ to correspond to $100/(1+z)\kpchi$
at all redshifts, as a tradeoff between the statistics and the 'closeness'. 
In the left panel of Figure \ref{fig:msat} we plot $\frac{1}{A_{\ast}}$ 
obtained from Equation (\ref{eq:invB}) as a function of satellite 
stellar mass $m_{\rm 2,s}$ at a median redshift of $z=0.3$. 
Four samples of central galaxies are shown by different colored lines, 
with stellar masses of $30.0$ (black), $15.0$ (red), $7.5$ (green) 
and $4.0$ (blue), respectively, in unit of $10^{10}\msunhi$. 
The corresponding halo masses lie in the range from $10^{12.4}\msunhi$ 
to $10^{13.1}\msunhi$.
Galaxy pairs are selected down to a stellar mass ratio of $1:32$, 
except for the lowest $m_{\rm 1,s}$ for which the lowest $M_{\rm 2,s}$ 
is not included, as galaxies at this mass consist of only $\sim 30$
particles, and are subject to resolution effect.
Error bars are plotted assuming Poisson fluctuations for merger pairs. 
The number of central galaxies being considered is 326, 505, 940 and 1275, 
respectively, as the primary mass is lowered. 
Close pair counts increase as the satellite mass decreases. 
For the $m_{\rm 1,s}=30.0\times 10^{10}\msunhi$ sample, the number of 
close companions for each central galaxy $N_{\rm p}(<100/(1+z)\kpchi)$ 
ranges from less than 2.0 for the most massive satellites, to 11.0 for 
the satellites with the lowest mass. There is a tendency for $\frac{1}{A_{\ast}}$ 
to decrease as the central stellar mass decreases;
$\frac{1}{A_{\ast}}$ changes by a factor of 3 for the  
range of $m_{\rm 1,s}$ being explored.

\begin{figure}
\begin{center}
\plotone{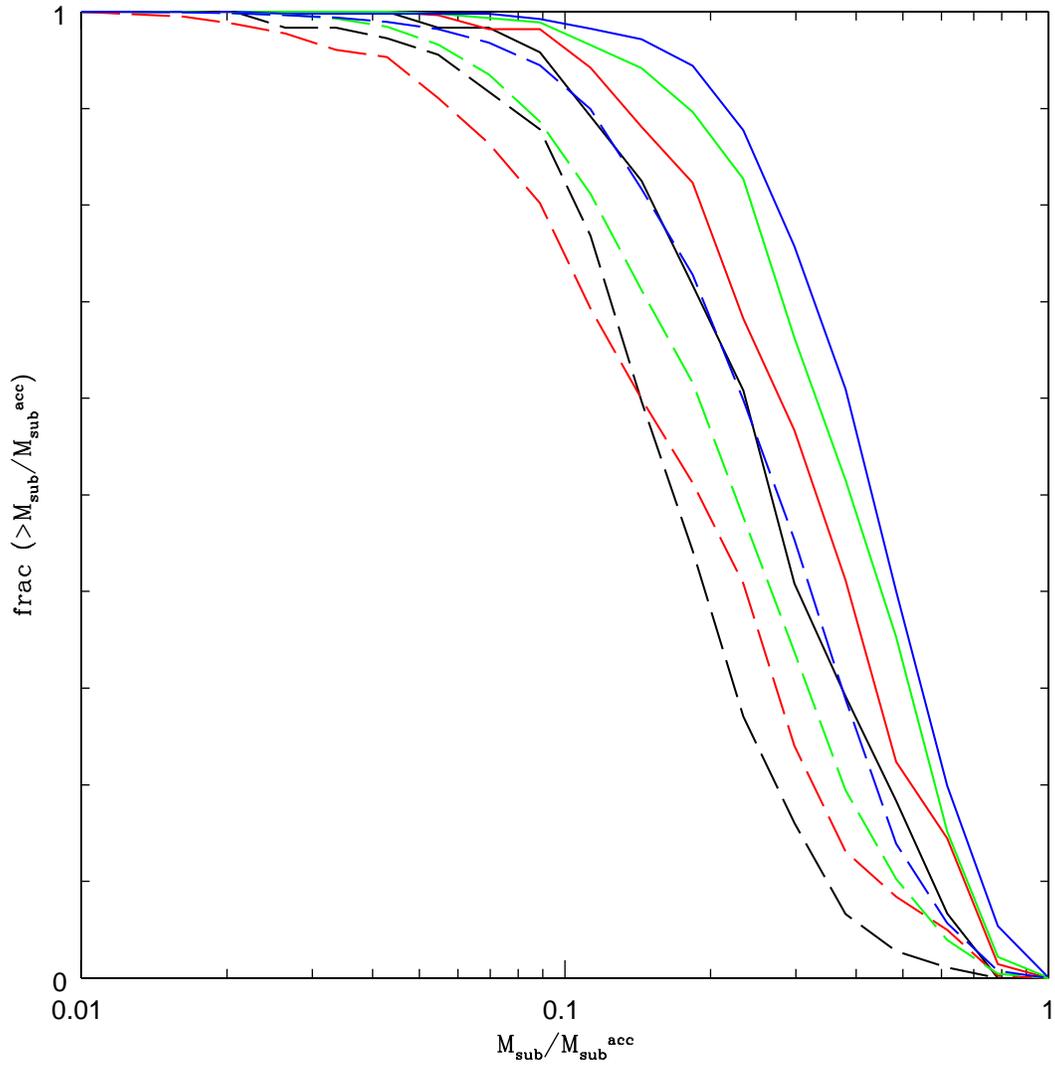}
\caption{\label{fig:subloss}
Accumulated distributions of retained mass fraction of subhalos
since accretion, for subhalos around $100/(1+z)\kpchi$ ($z=0.3$)
away from the halo center (solid lines), and those $10-60\kpchi$ away from
the halo center at $z=0.3$ (dashed lines).
The virial masses of primary halos are similar to those
hosting the central galaxies which are denoted with the same color
in the left panel. The stellar mass ratio of the satellite
inhabiting the subhalo to the central galaxy is above $1:30$.
}
\end{center}
\end{figure}

The fact that $\frac{1}{A_{\ast}}$ increases with  
$m_{\rm 1,s}$ implies that the stellar mass is not an optimal proxy
for $m_2$. It is not unexpected as the dynamical friction timescale is
determined by the total mass associated with the satellite. 
In this respect, subhalos are found and traced 
using a subhalo finder developed by \cite{han12}.
Here the method is extended to count in the baryonic particles. 
The right panel of Figure \ref{fig:msat} shows a similar 
plot as that in the left panel, except that the total satellite mass is   
taken for $m_{\rm s}$. As expected, $\frac{1}{A_{\ast}}$ remains approximately 
constant for each case of central-satellite pairs, which confirms that 
the total satellite mass is responsible for the merging process. In fact,
the behavior of  $\frac{1}{A_{\ast}}$ showed in the left panel of 
Figure \ref{fig:msat} can be well explained by Figure \ref{fig:subloss}, 
which displays the accumulated 
distribution of retained mass fraction of subhalos at $z=0.3$ 
since being accreted. $M_{\rm sub}$ denotes the total self-bound mass of 
subhalos, including dark matter, stars and gas particles. $M_{\rm sub}^{acc}$ 
is $M_{\rm sub}$ at infall. The stellar mass ratio of the satellite 
inhabiting the subhalo to the central galaxy is above $1:30$. 
The virial masses of primary halos are similar to those
hosting the central galaxies which are denoted with the same color 
in the left panel. Solid lines show results for subhalos at $100/(1+z)\kpchi$ 
away from the halo center (also at $z=0.3$) . 
We can see that above $70\%$ of subhalos still retain over $20\%$ 
of their original mass before accretion. The retained mass
is far more than the galaxy stellar mass residing in the center of 
the subhalo, given that the ratio of stellar mass to halo mass 
is at most a few percent \cite[see, e.g.,][]{behroozi12}.
Even at a distance of $10-60\kpchi$ which is widely used in 
literature, above $40\%$ of subhalos have a retained fraction of 
$\ge 20\%$ (the dashed lines). Thus, the total mass that is at work 
driving the merging outweighs the stellar mass of the galaxy. 

Figure \ref{fig:subloss} also shows that the fraction
of the mass (in terms of the subhalo mass at infall) retained by
the satellites of a fixed stellar mass $m_{2,s}$ increases with the
decrease of the host mass (from the black to the blue lines).
This is because the same $r_{\rm p}$ corresponds to 
a smaller fraction of the halo size for a more massive host halo, and  
thus satellites in larger host halos have gone deeper in the 
potential well and have lost a larger fraction of dark matter, 
as indicated in Figure \ref{fig:subloss}. Because of
the dependence of the total mass retained by the satellites on the
host mass, $\frac{1}{A_{\ast}}$ would show an increase with the
increase of the host mass if $m_{2,s}$ is used for $m_2$ in Equation
(\ref{eq:invB}).

\subsubsection{Accounting for Satellites' Mass Loss}

As discussed above, the total bound mass of the satellite (stellar mass 
$+$ retained dark matter mass) is what we need for $m_2$ in 
Equation (\ref{eq:invB}). However, it is not feasible to directly
measure the total mass for observed satellites in a large statistical 
sample. In order to obtain a convenient way of relating galaxy 
pair counts to merger rates, as a second try we replace 
$m_{2}$ with the total bound mass of the satellite. 

The total bound mass of a satellite evolves when it spirals into the
center of the host. The physical mechanism is the gravitational tidal
stripping that reduces its bound mass. There exist numerous works which
study the mass loss of subhalos \citep{tormen98,gao04,zentner05,diemand07,
giocoli08}. For the purpose of our current work, we want to know how 
the mass loss of a subhalo depends on its position in the host halo in 
a {\it statistical} way. \cite{han12} showed that the size of subhalos, 
on average, is proportional to the radius where the subhalo density 
equals the background density. Under the assumption of an isothermal 
sphere for the halo, we can easily deduce that the retained mass of 
the satellite is proportional to $r_{\rm p}$. 
Therefore, the retained bound mass can be modeled as 
$m_{\rm sub}^{acc}\frac{r_{\rm p}}{r_{\rm 1,v}}$. 
However, it is hard to estimate the infall information from observations. 
We thus have a different try, in which the total bound mass at infall 
$m_{\rm sub}^{acc}$ is taken as the virial mass $m_{\rm 2,v}$
that a central galaxy with the same stellar mass as the satellite
would have at the redshift being considered. Given that the relation
between the halo mass and stellar mass evolves slowly with
time \citep[see, e.g.,][]{wanglan10,behroozi12,yang12}, the two halo masses
do not differ much. Now the retained bound mass is modeled as 
$m_{\rm 2,v}\frac{r_{\rm p}}{r_{\rm 1,v}}$.

On the other hand, the spherical collapse model of halos indicates
$v_{\rm c} \propto H(z)r_{\rm 1,v}$, where $H(z)$ is the Hubble
parameter at redshift $z$. Thus, we have $m_{\rm 1,v} \propto {r_{\rm
    1,v}}^3\rho_{\rm crit}(z) \propto \frac{{v_{\rm c}}^3}{GH(z)}$, in
which $\rho_{\rm crit}(z)$ is the critical density at redshift
$z$. With this relation we have $v_{\rm c}=\sqrt{\frac{Gm_{\rm
      1,v}}{r_{\rm 1,v}}} \propto [m_{\rm 1,v}GH_{\rm
  0}E(z)]^{1/3}$. Replacing $m_2$ in Equation (\ref{eq:phi}) with the 
bound mass $m_{\rm 2,v}\frac{r_{\rm p}}{r_{\rm 1,v}}$ and using the 
above relations, we have
\begin{equation}
\label{eq:phi_re}
\Phi = B_{\ast} \frac{m_{\rm 2,v} 
n_{\rm 1}N_{\rm p}(<r_{\rm p}) [m_{\rm 1,v}GH_{\rm 0}E(z)]^{1/3}} {m_{\rm 1,v}r_{\rm p}}. \\
\end{equation}
where $H_{\rm 0}$ is the Hubble parameter at $z=0$, and 
$E(z)=\Omega_{\rm \Lambda}+\Omega_{\rm m}(1+z)^3$.
Figure \ref{fig:mcur} shows that, $\frac{1}{B_{\ast}}$ is 
nearly unchanged with respect to the primary galaxy mass and 
the satellite galaxy mass out to redshift $2$. This means that 
Equation (\ref{eq:phi_re}) is valid to connect the number 
of close pairs to galaxy merger rate. We fit the $\log \frac{1}{B_{\ast}}$ 
values for each primary galaxy to a straight line with zero slope. 
The four fitted values 
are then averaged to give a final result at each redshift. 
The resulting mean value of $\log\frac{1}{B_{\ast}}$ is $-0.23$ at $z=0.3$, 
$-0.24$ at $z=0.8$, $-0.25$ at $z=1.3$, and $-0.21$ at $z=2.0$, 
respectively, shown as the dotted lines in the plot. 
These values are very close, giving a mean $\frac{1}{B_{\ast}}$ of 
$10^{-0.23}$ with fluctuations among different redshifts less than $10\%$. 
This indicates that the redshift dependence can be well accounted for 
by Equation (\ref{eq:phi_re}). 

Although the radial extent $r_{\rm p}$ varies at different redshifts 
in Figure \ref{fig:mcur}, we explore the applicability of 
Equation (\ref{eq:phi_re}) when different radial extent is adopted at the 
same redshift. Figure \ref{fig:mcur_rc} shows results 
with $r_{\rm p}=50/(1+z)\kpchi$ and $150/(1+z)\kpchi$ at $z=0.3$.
In both cases, $\frac{1}{B_{\ast}}$ is consistent with that obtained 
using $r_{\rm p}=100/(1+z)\kpchi$ (dotted lines).

The above results show that Equation (\ref{eq:phi_re}) can accurately
describe the merger rate (at an accuracy $\sim 10\%$). This
indicates that the average merger timescale for a pair of galaxies at
a small separation $r_{\rm p}< 150 \kpchi$ is
\begin{equation}
\label{eq:Tmg_re}
  T_{\rm mg} =10^{-0.23} \frac{m_{\rm 1,v}}{m_{\rm 2,v}} [m_{\rm 1,v}GH_{\rm 0}E(z)]^{-1/3}r_{\rm p}, 
\end{equation}
where the coefficient $\frac{1}{B_\ast}$ is already replaced by its 
mean value $10^{-0.23}$. The timescale linearly scales with the (physical) 
separation, and depends on the redshift through $E^{-1/3}(z)$. 
Its strong dependence on
the halo masses $m_{\rm 1,v}$ and $m_{\rm 2,v}$ tells us that one must
take into account the mass dependence when measuring the merger rate
from the count of close pairs. Even for major mergers (which are
routinely defined as the mergers with the stellar mass ratio above 
$1/4$), the timescale could change significantly. 

\subsection{Satellite-Satellite Pairs}
\label{sec:sat-sat}

\begin{figure}
\begin{center}
\plotone{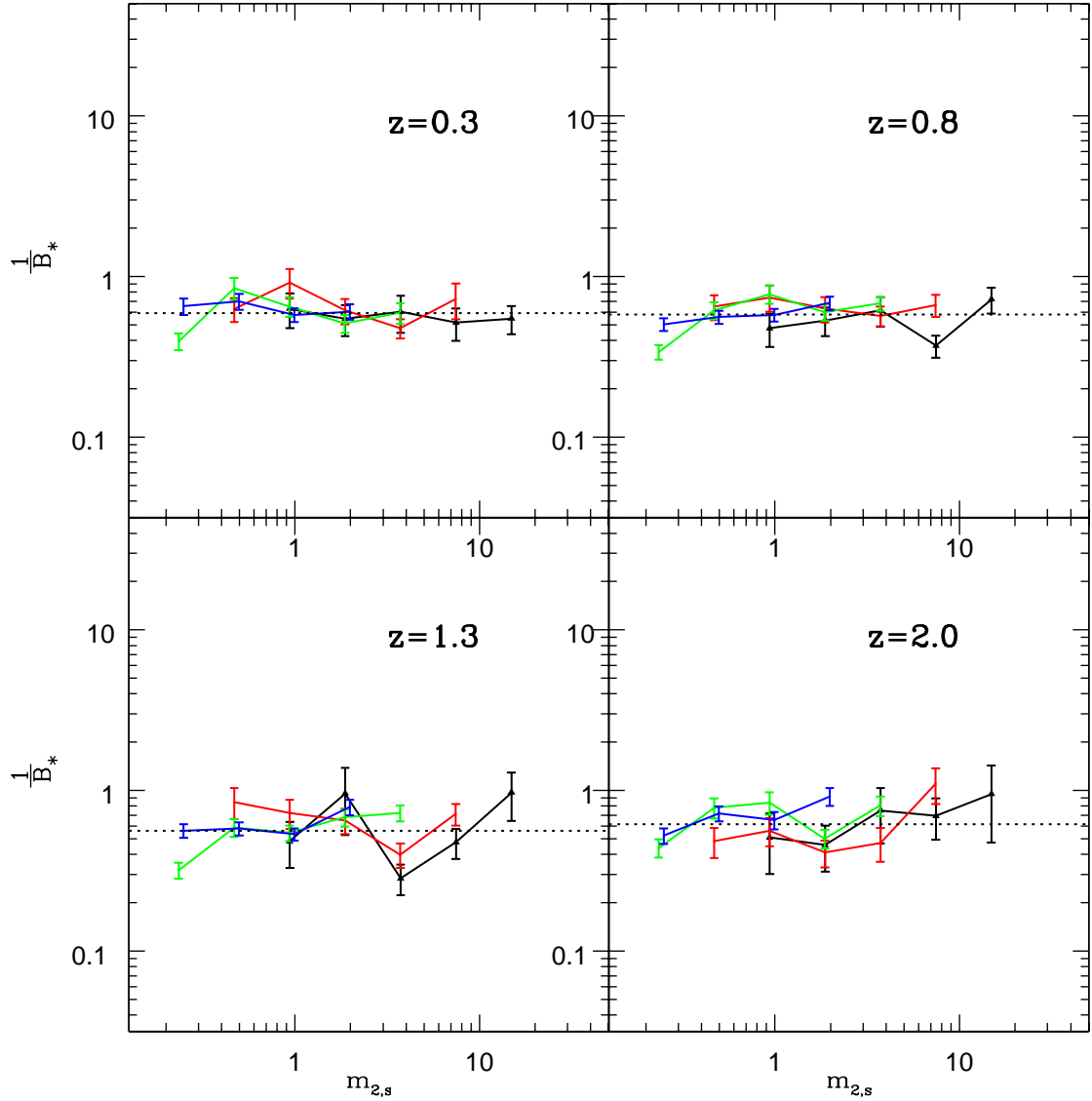}
\caption{\label{fig:mcur}
Same as the left panel of Figure \ref{fig:msat}, but the virial mass for 
satellites is used (Equation \ref{eq:phi_re}). Results are shown in 
4 redshifts as displayed in different panels. 
The dotted lines are the fitting results at each redshift, 
showing $\log10(\frac{1}{B_{\ast}})=-0.23,-0.24,-0.25,-0.21$ respectively, 
as the redshift increases. 
}
\end{center}
\end{figure}

\begin{figure}
\begin{center}
\plotone{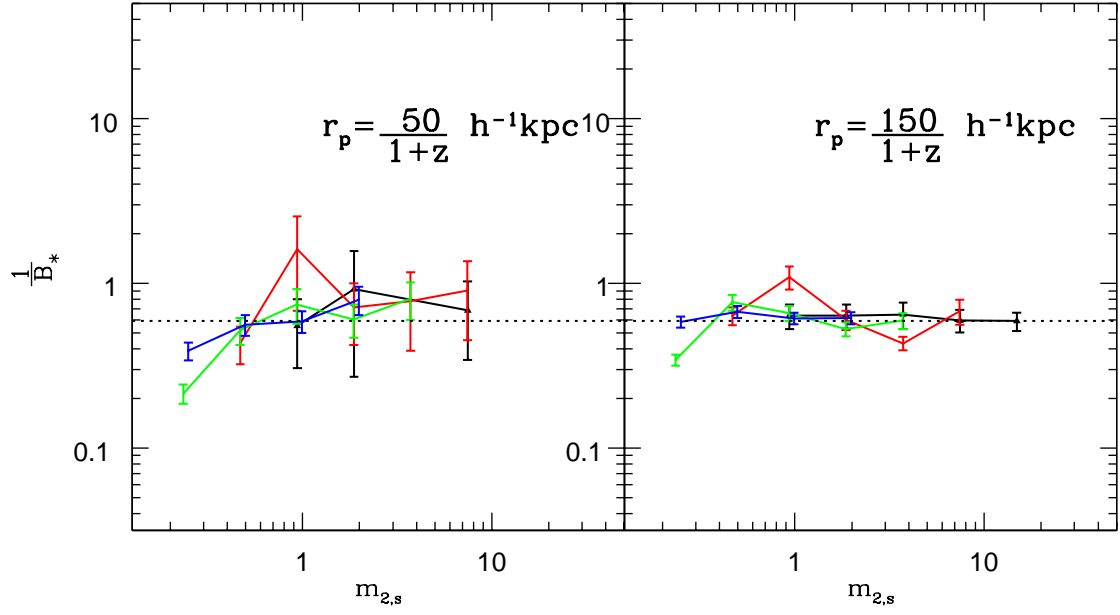}
\caption{\label{fig:mcur_rc}
Same as the upper left panel in Figure \ref{fig:mcur}, but the outer
radius corresponds to $50/(1+z)\kpchi$ (left), and
$150(1+z)\kpchi$ (right).
}
\end{center}
\end{figure}

\begin{figure}
\begin{center}
\plotone{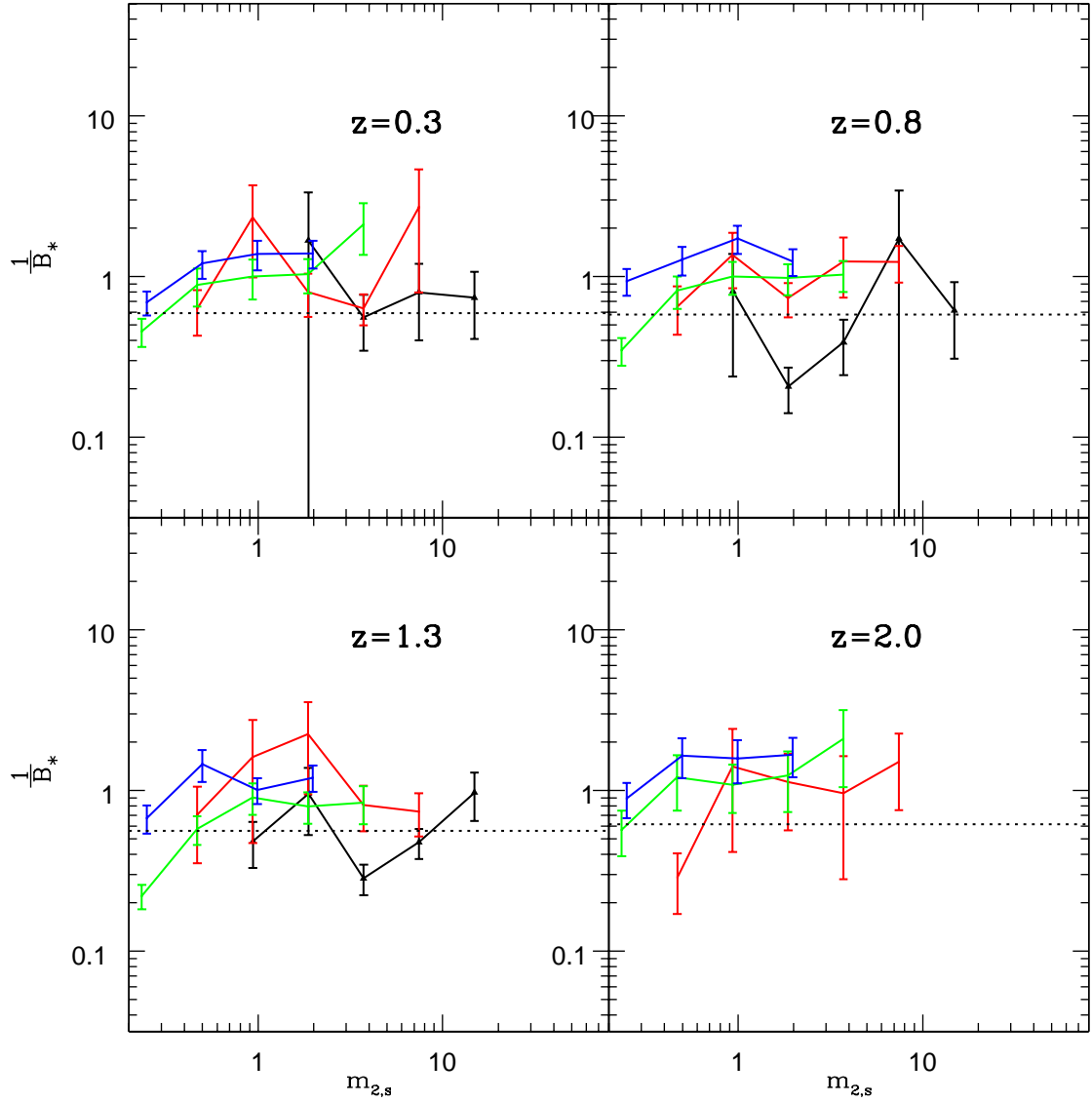}
\caption{\label{fig:mcur_sat}
Same as Figure \ref{fig:mcur}, but for satellite-satellite pairs.
The dotted lines are the same as in Figure \ref{fig:mcur}.
}
\end{center}
\end{figure}

In section \ref{sec:cen-sat}, we have obtained a good relationship 
between the merger rate and close number counts within $100\kpchi$ 
from central galaxies. We have considered the central-satellite 
pair which is the dominant merging pattern. However, the 
satellite-satellite merging pairs also exist, although they are 
minor in comparison to central-satellite pairs. We check in Figure  
\ref{fig:mcur_sat} whether the scaling relation we have obtained 
still holds for satellite-satellite merging pairs. Dotted lines 
in each panel are the same as in Figure \ref{fig:mcur}. At each 
redshift, most part of the four lines slightly lies above the 
dotted line. After applying the same fitting method as in Figure  
\ref{fig:mcur} to the data, the resulting amplitude is typically about 
$0.15$ dex higher than those for central-satellite pairs in Figure  
\ref{fig:mcur}. This higher amplitude results from the
lower merger rate for satellite-satellite pairs, compared to 
that for central-satellite pairs with the same masses \citep{jiang10}.
At small separation ($\le 100\kpchi$), the contribution to pair counts 
from satellite-satellite pairs is at the level of $10\%$ 
\citep[see, e.g.,][]{yoo06}. In view of the small increase in 
$\frac{1}{B_{\ast}}$ for satellite-satellite pairs and the low 
fraction of such pairs, the influence of satellite-satellite pairs on 
the relation of merger rates and pair counts is expected to be 
very weak. We can therefore adopt the relation established only 
with central-satellite pairs.

\subsection{Application to observations}
\label{sec:apl_obs}

So far, all the investigations were done in the 3-dimensional
frame. In observations, what we usually get is the number count of
galaxy pairs within a projected separation $r_p$ (e.g. from the
correlation function measurement). Some of the close pairs in
projection are foreground or background pairs which are separated
beyond $r_p$ in the real 3-dimensional space.  To account for the
projection effect, one needs to introduce a correction factor $f_{\rm
  3d}$ to convert a projected pair count ${N_{\rm p}}^{proj}$ to a
3-dimensional count, $N_{\rm p}=f_{\rm 3d}{N_{\rm p}}^{proj}$.

In \cite{jiang12}, they found that the projected number density 
of satellites obeys a power-law form with the best-fit 
logarithmic slope of $-1.05$, and this profile is independent 
of both central galaxy luminosities and satellite luminosities. 
According to their Equation (8),  
for projected number of satellites within $r_{\rm p}$ 
which is equivalent to number of pairs in the central-satellite 
case, it has a fraction $f_{\rm 3d}=66\%$ within the real 3-dimensional 
$r_{\rm p}$. 
Halo masses in their sample are above $10^{12}\msun$, spanning a wider 
range than the primary halo masses being considered here. 
The power-law slope for the satellite radial distribution 
is fixed out to the virial radius ($>300\kpc$). 
Therefore, this 3-dimensional fraction $f_{\rm 3d}$ is the 
same for all close pair cases ($<100\kpchi$).
For satellite-satellite pairs, we use the simulation to check 
whether $f_{\rm 3d}=0.66$ is also applicable or not. 
We project all galaxies in the simulation box on the
$x-y$ plane while assuming the $z$-direction is along the line-of-sight,
which is equivalent to selecting companions in a
photometric survey with $\frac{\Delta z}{1+z}\le 0.015$.
We find that $f_{\rm 3d}=0.66$ are generally consistent 
with the simulation results, and thus can also be applied to 
satellite-satellite pairs.

According to $\Phi =n_{1}N_{\rm p}(<r_{\rm p})/T_{\rm mg}$, 
when the 3-dimensional pair count per unit volume $n_{1}N_{\rm p}(<r_{\rm
  p})$ is known, we can get the merger rate per unit volume by using 
Equation (\ref{eq:Tmg_re}) for the merger timescale. But for 
observations, it is the pair count in projection $N_{\rm p}^{proj}$ 
that is more easily accessible.  
If $N_{\rm p}^{proj}$ is used to measure the merger rate,
i.e. $\Phi =n_{1}N_{\rm p}^{proj} (<r_{\rm p})/T_{\rm mg}^{proj}$, 
the projection effect must be taken into account and 
the merger timescale becomes  
\begin{equation}
\label{eq:Tmg_fin}
T_{\rm mg}^{proj}(<r_{\rm p})  = \frac{10^{-0.23}}{0.66} \frac {m_{\rm 1,v}}{m_{\rm 2,v}} 
[m_{\rm 1,v}GH_{\rm 0}E(z)]^{-1/3} r_{\rm p}. \\
\end{equation}

Equation (\ref{eq:Tmg_fin}) offers us a method of computing the 
timescale for mergers between primary galaxies with stellar mass 
$m_{\rm 1,s}$ and secondary galaxies with stellar mass 
$m_{\rm 2,s}$ within a projected radius of $r_{\rm p}$. 
Besides the radial extent $r_{\rm p}$ and redshift $z$, 
there are two other input quantities in Equation (\ref{eq:Tmg_fin}): 
$m_{\rm 1,v}$ and $m_{\rm 2,v}$.
$m_{\rm 1,v}$ is the median virial mass of isolated halos which host 
central galaxies of $m_{\rm 1,s}$ at redshift $z$. 
Similarly, $m_{\rm 2,v}$ is the median virial mass of isolated 
halos which host central galaxies of $m_{\rm 2,s}$ 
at redshift $z$. These virial masses, which are defined in terms of $200$ 
times the critical density, can be estimated from the 
relation between dark halo mass and central galaxy mass 
\citep[e.g.][]{yang03,tinker05,mandelbaum06,wanglan06,yang07,zheng07,
guoqi10,moster10,wanglan10,more11,li12,behroozi12,yang12,velander13}.
The error of measuring the merger rate through Equation (\ref{eq:Tmg_fin}) 
comes from the uncertainty of the close pair count and from the accuracy 
of the merger time scale. We estimate that the accuracy of our formula 
is at $10\%$ level. To reach a similar level of the merger rate measurement, 
one would need at least a few hundred of close pairs.

\section{Discussions}

\begin{figure}
\begin{center}
\plotone{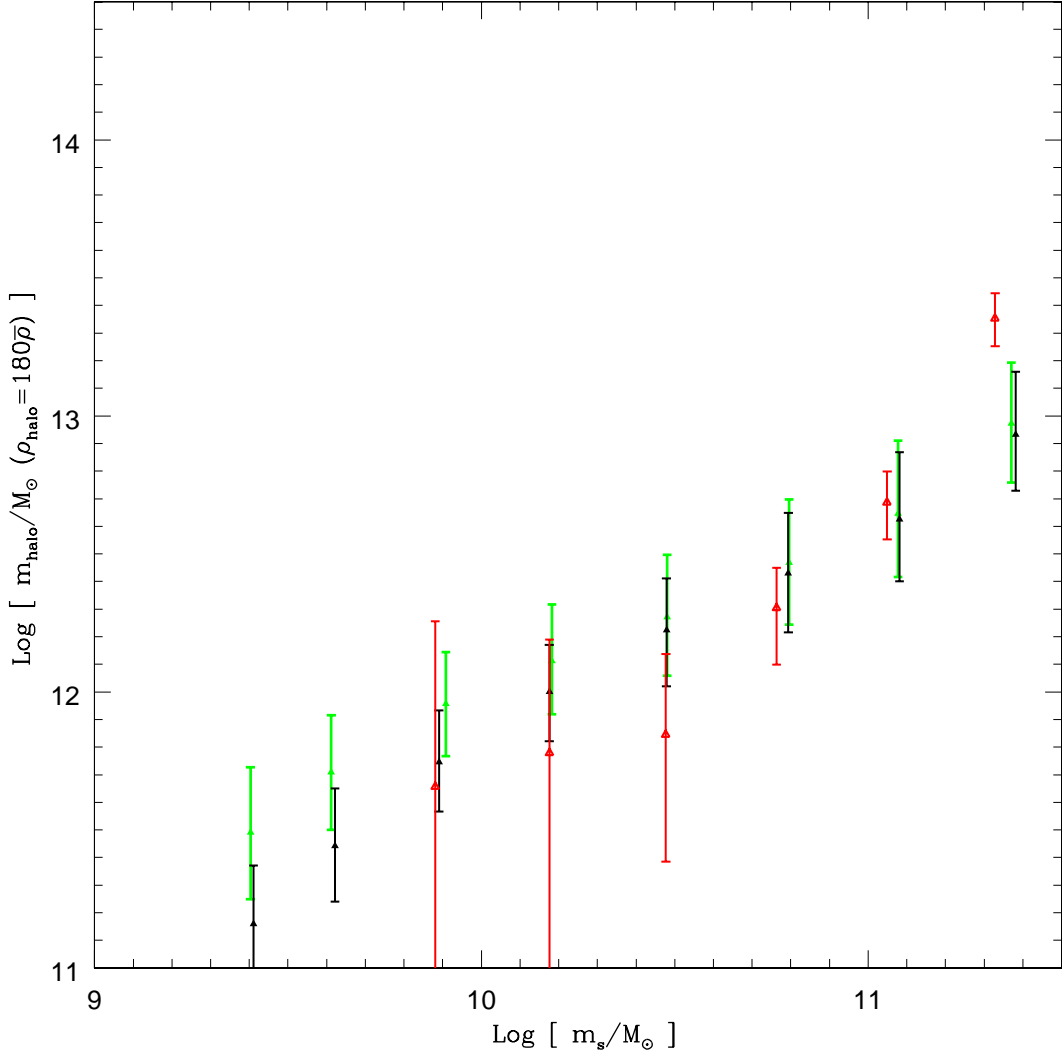}
\caption{\label{fig:ms_mv}
Median halo mass as a function of its central stellar mass at $z=0.3$
(points with error bars showing $68.3\%$ interval), black for the 
default galaxy identification method (b=0.025) and green for b=0.0125.  
The bin width of the stellar mass is $0.3$ in logarithmic space.
To compare with the results from \cite{mandelbaum06} (red points),
we define halo mass here as that enclosed within the radius, where
the overdensity is equal to 180 times the mean density.
}
\end{center}
\end{figure}

In this section we will first discuss about the effect of modeling 
uncertainties of galaxies in the simulation. As we know, the baryonic 
physics governing galaxy formation and evolution is not
fully understood yet, though the simulated galaxies already resemble 
closely those in the real Universe.  As we have shown, the merger 
timescale depends on the masses of the satellite and the host halo,
therefore it is important to check the ratio of the stellar mass to the 
dark matter halo mass of a galaxy, to see if the simulation is
suitable for the study of the merger rate of galaxies. 
We show in Figure \ref{fig:ms_mv} the relation between the halo 
mass and stellar mass of the central galaxy for the lowest redshift 
case. Results are compared to those observed by  
\cite{mandelbaum06}. It is clear that they are generally consistent, 
except for the highest mass bin where the halo mass is about $2.5$ 
times smaller in our simulation. This can be attributed to the 
gas overcooling in massive halos, which results from the lack 
of AGN feedback to a certain degree. Since most massive galaxies 
are central ones, the overcooling mainly affects the 
identification of merger remnants, while having little effect on 
the descending of satellites. As we emphasized in the end of \S
\ref{sec:apl_obs}, if we can find out the host halo mass of the
central galaxies from observations (say from the halo occupation
distribution method), we can compute the merger timescale accurately.

\begin{figure}
\begin{center}
\plotone{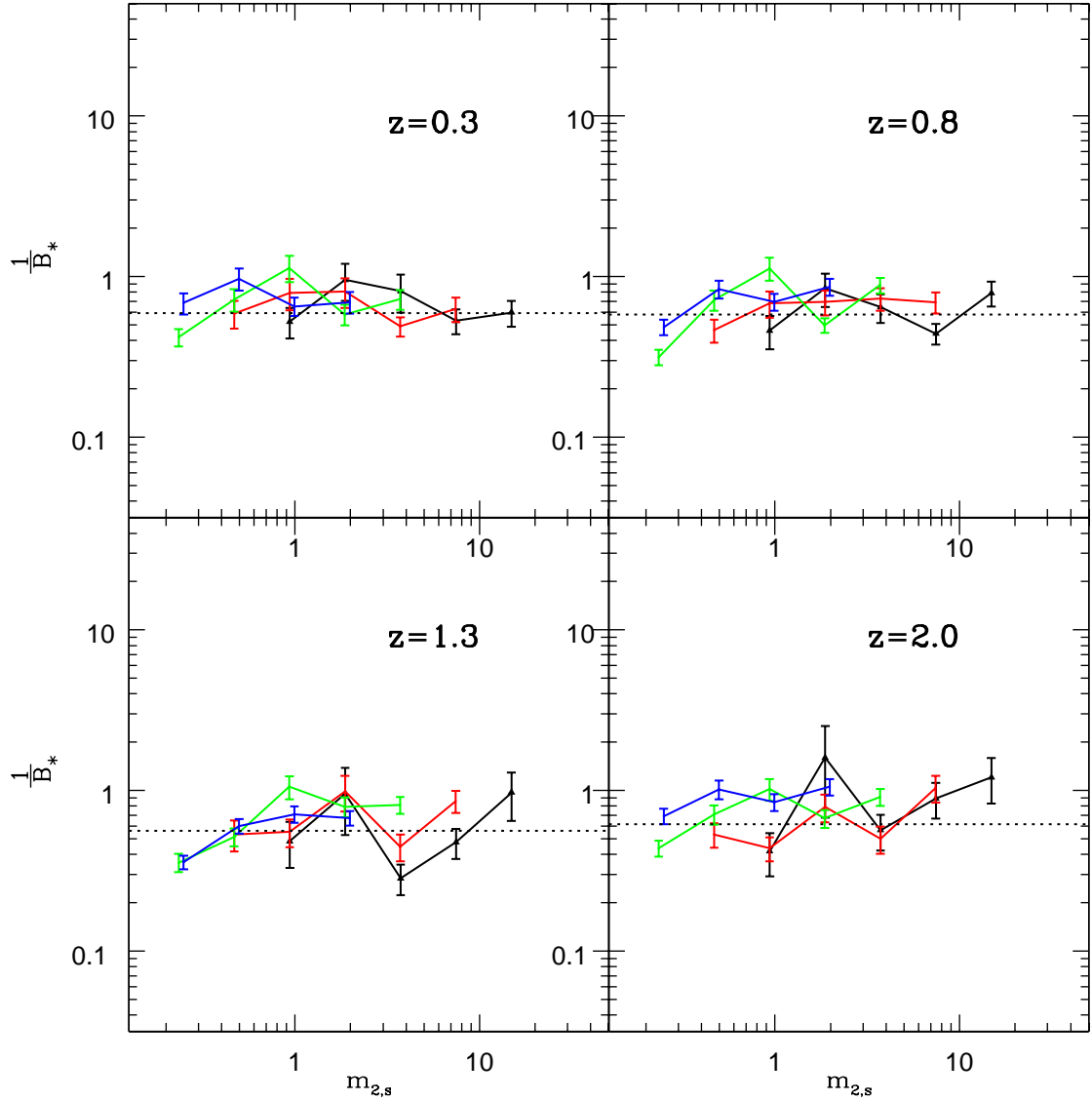}
\caption{\label{fig:mcur_b01}
Same as Figure \ref{fig:mcur}, but galaxies are linked
with a smaller linking length ($b=0.0125$). Dotted lines are the same 
as those in Figure \ref{fig:mcur}.
}
\end{center}
\end{figure}

The gas overcooling problem of the most massive galaxies in the
simulation should have negligible effect on our results for the merger
time scales, because we do not explicitly use the stellar mass in the
equations.
This is further illustrated with a different identification method 
which will leads to a different correspondence between the halo mass and 
the galaxy stellar mass (green points with error bars in Figure 
\ref{fig:ms_mv} ). Here we use a smaller linking length ($b=0.0125$) 
when identifying galaxies. For a given host halo mass, the galaxy stellar
mass is smaller than the galaxies identified with $b=0.025$ in the
previous section. Because we use the correct dark halo mass for each
galaxy in Equation (\ref{eq:phi_re}), we would expect very little
change from the case of $b=0.025$. This is shown in Figure  
\ref{fig:mcur_b01}, which shows that the change is indeed small when 
the new results are compared to the dotted lines which
are the same as in Figure \ref{fig:mcur}. The largest change happens
at $z=2.0$, where $\frac{1}{B_{\ast}}$ is higher than that in Figure  
\ref{fig:mcur} by about $20\%$. Only the central core of a galaxy
would be selected into a galaxy again after shortening the linking
length. It makes galaxies more easily stripped of particles, thus
delaying the merger identification or reducing identified merger
events. This explains why $\frac{1}{B_{\ast}}$ is relatively higher
here.  The results reinforce the conclusion that the merger timescale
is dominated by the gravitational process, as shown by
\cite{jiang10} who found that even if the baryonic mass is halved in the
simulation, the time needed for a satellite at the host virial radius to 
merge with the central galaxy generally changes at the level
of $10\%$. We therefore expect that the merger timescale for close
galaxy pairs would change at a similar level if the stellar mass
changes by less than $2$ times.
 
Next we make a comparison with results from \cite{kitzbichler08}. 
We calculate $T_{\rm mg}^{proj}$ for galaxies more 
massive than $4\times10^{10}\msunhi$ within a projected 
distance of $100/(1+z)\kpchi$ at $z=0.1$.                      
\cite{kitzbichler08} gives a timescale of $9.0$ Gyr for such 
mergers. Since their stellar mass ratio is between 1:1 and 1:4, 
we compute $T_{\rm mg}^{proj}$ accordingly, finding $T_{\rm mg}^{proj}$ 
is between $1.6$ Gyr and $8.2$ Gyr ($h=0.71$), corresponding to 1:1 
mergers and 1:4 mergers respectively. The stellar mass is 
converted to halo mass using the relation between halo mass 
and central stellar mass established in \cite{mandelbaum06}. 
The timescale computed according to \cite{kitzbichler08} ($9.0$ Gyr) is 
only close to that for 1:4 mergers in our results.
On the other hand, the merger timescales for the same galaxies 
but within projected $20\kpchi$ are between $0.31$ Gyr (1:1) and 
$1.6$ Gyr (1:4), 
by assuming $T_{\rm mg}^{proj}$ is proportional to the outmost radius.
These all indicate that a single timescale 
is not sufficient to describe mergers with stellar mass ratios in the 
range from 1:1 to 1:4. We need to consider the dependence 
of the timescale on both virial masses of the two galaxies.

\section{Conclusions}

Using a high-resolution N-body/SPH cosmological simulation, we have
studied the relation between galaxy pair counts and merger rates.  Our
results indicate that galaxy merger rates can be inferred at an
accuracy $\sim 10\%$ from the number of close pairs as shown by
Equation (\ref{eq:phi_re}). The equation gives explicit dependence on
the bound masses of galaxies, on the pair separation and on the
redshift. Equivalently, the merger timescale can be established for
galaxy pairs with the physical separation less than $150\kpchi$ down to a
stellar mass ratio of $1:30$ out to redshift $2.0$.

Although stellar mass can be directly measured in galaxy surveys, the
dynamical friction time scale is determined by the total bound mass
surrounding the galaxies. By using the virial mass of both galaxies in
the close pair and properly considering the mass loss of subhalos, we
present an accurate scaling relation (Equation \ref{eq:Tmg_fin}) for
galaxy mergers within a projected distance $r_{\rm p}$. For this
equation to be applied to real observations, the virial mass of the
galaxies should be obtained with the techniques such as the halo
occupation distribution method. The additional advantage of using the
bound mass for the merger timescale is that our results are rather
insensitive to the modeling details of galaxies in the simulation.

For major mergers (1:1-1:4) of galaxies above $4\times10^{10}\msunhi$
at $z=0.1$, estimated merger timescales ($r_{\rm p}<100\kpchi$) based
on Equation (\ref{eq:Tmg_fin}) spans a wide range from $1.6$ Gyr to
$8.2$ Gyr ($h=0.71$), corresponding to 1:1 mergers and 1:4 mergers
respectively.  In contrast, the timescale given by
\cite{kitzbichler08} ($9.0$ Gyr) is only close to that for 1:4 mergers
in our results.  Our results indicate that a single timescale is not
enough to describe mergers of stellar mass ratios even only from 1:1
to 1:4, and the dependence on the virial masses of the galaxies must
be taken into account.

Although all the equations are formulated for central-satellite pairs,
satellite-satellite merging can also be approximately described, with
a typical deviation of $0.15$ dex for the normalizing factor. This is
a consequence of relatively lower merger rates for satellite-satellite
pairs. Our results for the satellite-satellite can be incorporated
into an accurate modeling of the merger rate if the satellite fraction of
galaxies is known (say, from HOD modeling \citep[e.g.]{zheng07}).
Since the fraction of satellite-satellite pairs is low (at $\sim 10\%$
level), the influence of such pairs to the relation of merger rates
and pair counts is expected to be very weak. The relation established
with central-satellite pairs can therefore be applied to all galaxies
pairs to a good approximation.

\acknowledgments
This work is  sponsored  by NSFC  (11320101002，11121062, 11033006,
11003035) and the CAS/SAFEA International Partnership Program for  Creative 
Research Teams  (KJCX2-YW-T23). The simulation was performed at the
Shanghai Supercomputer Center. Part of the data analysis was done on 
the supercomputing platform at Shanghai Astronomical Observatory.

\end{document}